# An Advice about Shimming in High-Resolution Nuclear Magnetic Resonance

V.V Korostelev

*Kaliningrad University, Kaliningrad, Russia*

## Abstract

Three methods of active shimming in high-resolution NMR in existence (manual shimming, lock optimization and gradient shimming) are briefly discussed and their advantages and shortcomings are compared and an advice on their use is given.

There are two main approaches[1] to active shimming in modern NMR: manual shimming and auto-shimming. The auto-shimming techniques include lock optimization and gradient shimming. There are two methods of gradient shimming: 1D *z*-gradient shimming and 3D gradient shimming. All of these techniques use Golay coils (*1*), which are also called shims nowadays. This paper highlights advantages and shortcomings of these techniques with stress on the latest developments in 3D gradient shimming. This also explains a relative value of these techniques and helps to decide which kind of shimming technique is most appropriate to apply in each particular case and why.

Traditionally, shimming is performed manually, which has many limitations. This does not always yield satisfactory results since the lock signal normally used in manual shimming and lock optimization does not show how static magnetic field in-homogeneity is distributed in space and, therefore does not answer the question – which shim needs a correction? Also, manual shimming makes on user an impression of total control over process, where s/he makes intelligent guess on what shims to correct. This psychological aspect of manual shimming is very different from automated techniques where software decides on what shims to correct and, therefore, role of a user becomes less active. However, such active role of user as in manual shimming quickly becomes tedious since this distracts from a more interesting NMR experiment, which normally follows shimming procedure, and more importantly, manual shimming can be a difficult task.

Experience of several generations of NMR spectroscopists suggests that good shims normally can be found for particular magnet, probe and sample, as a result of manual shimming during less than an hour[2]. Although an experienced user normally does manual shimming much faster than that (since this skill can be improved with experience), a novice may experience real difficulties and even get stuck using manual shimming. Moreover, the question whether the minimum of the static magnetic field achieved is local or global cannot be answered after manual shimming experiment because 3D distribution of the static magnetic field is still unknown. Practically, probability of getting into local minimum after manual shimming is very high.

Another method of active shimming, lock optimization (*2*) was first auto-shimming technique introduced before gradient shimming was implemented. This uses level of lock signal to estimate static magnetic field homogeneity and like manual shimming does not provide information about 3D distribution of the field.

---

[1] This does not include passive shimming, which historically was the first shimming technique and not in use in modern NMR. Early NMR magnets were permanent, with pole pieces produced from high quality steel and polished to optical flatness in order to improve the homogeneity of the static magnetic field. To obtain high field homogeneity, the pole faces needed to be parallel to each other. Thus, small pieces of thin sheet steel, called shims, were placed between the pole faces and the steel core of the magnet for adjustment of their position relative to each other. This method of shimming, later called passive shimming, was relatively rough and far from easy for the user.
[2] for high field magnets, for example, manual shimming started from all shims set to zeroes can take many hours.

In order to avoid this limitation of the manual shimming and lock optimization techniques, a more intelligent approach, which includes measurement of spatial static magnetic field distribution, was proposed. This approach is called automated gradient shimming and includes 1D *z*-gradient shimming and 3D gradient shimming techniques. Each of these techniques relies on measurement of static magnetic field in-homogeneity (either 1D or 3D, as it is suggested by their names) followed by numerical calculation of new shims, and this procedure is normally repeated several times, called iterations, automatically until satisfactory static field homogeneity is achieved. Such approach makes more sense than manual shimming since it is not anymore like blind walking[3] in shim and field spaces but a sensible shim adjustment based on measurement. Another feature of automated gradient shimming techniques is that these can be performed using either $^1H$ or $^2H$ meanwhile manual shimming and lock optimization use lock-channel, which is normally used for observation of $^2H$ only.

A most simple automated gradient shimming technique is called 1D *z*-gradient shimming and this corrects *z*-shims (*z1*, *z2*, *z3*, *z4*, *z5*) only. As other shims are not corrected during this procedure, it is less effective than shimming of all shims during even manual shimming, for example. An advantage of this technique is that correction of *z*-shims brings most significant improvement in field homogeneity since more spins affected by field in-homogeneity are distributed along *z*-axis than along *x*- and *y*-axis within a sample volume. Hence, 1D *z*-gradient shimming is useful for quick shimming starting from poor shims. Also a combination of 1D *z*-gradient shimming with manual shimming (latter – for non-*z* shimming) is often used in high-resolution NMR experiments.

A 3D expansion of gradient shimming is called 3D gradient shimming technique and this performs correction of all shims, not only *z* shims as in 1D *z*-gradient shimming. However, this technique is more time consuming than 1D *z*-gradient shimming and, in most cases, than manual shimming as well. There are two advantages of 3D auto-shimming to be mentioned here: first is that this technique normally offers better performance than 1D *z*-gradient shimming, field optimization and manual shimming as well, and, second, is that this technique normally requires less involvement of user into shimming. However, this requires from a user a certain familiarity with setup parameters and intrinsic detail of the technique, and therefore, it can be more difficult to start with for a novice. Another shortcoming of 3D gradient shimming is that this is highly dependent on correct working of hardware, in particular, on reproducibility of gradients (in particular, shim gradients since shim coils are not designed for purposes of NMR imaging).

---

[3] in fact, manual shimming uses some rough ideas about what kind of field in-homogeneity dominates in sample volume and these ideas can be drawn from appearance of line-shape of a spectrum.

One of 3D auto-shimming techniques, reported by van Zijl (*3*) in 1994 is based on 3D Fourier imaging (*4*) (in particular, on a field mapping[4] technique proposed by Maudsley (*5*) in 1979) and subsequent numerical calculations of the shim corrections. In general, field mapping can be performed using either shim gradients or gradients produced by PFG (Pulsed Field Gradient) modules – the both approaches can be valid and effective although some potential advantages of using PFG are sometimes suggested. One of the recent developments in 3D auto-shimming was a technique using shim gradients, which can be produced with every commercial NMR spectrometer equipped with shim coils and homo-spoil facility (*6,7*).

## Making choice of a shimming technique

From user's point of view it is important to make intelligent choice among available shimming techniques to perform before starting an NMR experiment. This requires estimation of the static magnetic field in-homogeneity within a sample volume, and the good starting point here is to measure line-width of a $^1$H spectrum acquired using line-shape test sample. If line-width is about one Hertz or less then the field is well shimmed and the current shims are good. If the line-width is about several Hertz or more (with all shims set to zero it can be about several hundred Hertz) then the field homogeneity and current shims are poor.

Although, manual shimming and lock optimization seem more familiar for many, these techniques may lead to local minima (as it was already mentioned above). Also, manual shimming is often tedious when started from poor shims. However, when shims are already good and only fine adjustment of shims (typically, high-order shims) is required, it's sensible to retreat to manual shimming.

When field homogeneity is poor then either 1D or 3D gradient shimming has to be used, depending on what one tries to achieve. Although 1D *z*-gradient shimming is able to give a quick and significant improvement of field homogeneity, it's unlikely to be the best possible result since transverse shims are not adjusted during this procedure and $^1$H line broadening after 1D *z*-gradient shimming is typically about, at least, several Hertz.

Hence, 3D gradient shimming, which optimizes all shims, is most sensible choice for simultaneous and iterative rough and fine adjustment of all shims when it starts from not very good shims. This also might be used for fine shimming starting from some very good shims but this is hardly sensible as 3D gradient shimming is time-consuming procedure and a faster result for fine shimming of few shims could be achieved with some manual adjustment instead of an automated one. However a user should be aware of the known shortcomings of 3D gradient shimming, which are its complexity, dependence on reproducibility of gradients and time to be spent. In general, 3D gradient shimming is the most time consuming procedure among all shimming techniques in existence and this normally takes no less than an hour.

---
[4] field mapping is a procedure for measurement of 3D static magnetic field in-homogeneity.

A sensible alternative to 3D gradient shimming can be a combination of 1D $z$-gradient shimming followed by manual shimming. This allows quicker compensation of field in-homogeneity by adjustment of $z$- and non-$z$ shims as well than 3D gradient shimming. The shortcomings of such combination are the same as for manual shimming, i.e. it can lead to a local minimum and this will require from a user an intelligent guess on what shims require manual adjustment, whatever this guess might be – right or wrong.

ACKNOWLEDGEMENTS

I thank Professor Gareth Morris from University of Manchester, UK for reading the draft of this paper and his suggestions how this can be improved. I tried my best doing this.